\documentclass[a4paper]{jpconf}

\usepackage{amsmath,amsthm,amssymb,epsf}
\usepackage{mathbbol,bbm}
\usepackage{graphicx}
\usepackage{wasysym}
\usepackage{citesort}
\usepackage{subfigure}

\newcommand{\beqa}{\begin{eqnarray}}
\newcommand{\eeqa}{\end{eqnarray}}
\newcommand{\f}{\begin{equation}}
\newcommand{\ff}{\end{equation}}

\newcommand{\bean}{\begin{eqnarray*}}
\newcommand{\eean}{\end{eqnarray*}}
\newcommand{\ra}{\rightarrow}

\def\be{\begin{equation}} \def\ee{\end{equation}}

\begin{document}

\title{Matter in Toy Dynamical Geometries}
\author{Tomasz Konopka}
\address{ITP, Utrecht University,Utrecht 3584 CE, the Netherlands}
\ead{t.j.konopka@uu.nl}

\begin{abstract}
One of the objectives of theories describing quantum dynamical
geometry is to compute expectation values of geometrical
observables. The results of such computations can be affected by
whether or not matter is taken into account. It is thus important
to understand to what extent and to what effect matter can affect
dynamical geometries. Using a simple model, it is shown that
matter can effectively mold a geometry into an isotropic
configuration. Implications for ``atomistic'' models of quantum
geometry are briefly discussed.
\end{abstract}

%Preprint numbers: ITP-UU-08/13, SPIN-08/12;

\section{Introduction}

In the context of quantum gravity (see e.g.
\cite{DynamicalGeometry} for an overview of different approaches),
much attention is devoted to studies of pure-gravity systems in
which all matter degrees of freedom and their interactions are
switched off. Whereas the use of this assumption is understandable
given that most proposals for quantum theory of gravity are
difficult to study even without the inclusion of additional
fields, it is known that the presence of matter can alter the
behavior of a gravitational system \cite{BirrellDavies}. The
purpose of this paper is to illustrate the effect of matter on
geometry in a simple setting. In particular, the purpose is to
show that matter can induce isotropy in a model that does not
assume it a-priori - an effect not easily seen from standard
treatments of matter in dynamical geometries \cite{BirrellDavies}.

Physical theories are often formulated using the Feynman path
integral and many candidates for a theory of quantum gravity
\cite{DynamicalGeometry} are also defined within this framework.
These candidates propose to study path integrals (partition
functions) formally written as \be \label{Zheuristic} Z_\phi =
\int \mathcal{D}[g] \int \mathcal{D}[\phi] \; e^{-E(g,\,\phi)} \ee
where integrals denote summations over geometries $g$ and matter
degrees of freedom $\phi$, and $E(g,\phi)$ is a weighting function
that depends on both the geometry and the matter content (this
function can be interpreted as the action or the energy of the
system, according to the setting). The weighting function is often
assumed to be of the form $ E(g,\,\phi) = E_g(g) + E_\phi
(g,\,\phi)$ so that the contribution from the matter fields in
$E_\phi(g,\,\phi)$ is separated from the pure-geometry part
$E_g(g)$. Furthermore, it is often assumed that if the matter
contribution $E_\phi(g,\,\phi)$ is set to zero, the resultant,
simpler, path integral \be \label{Zheuristic2} Z_{\phi=0} = \int
\mathcal{D}[g] \; e^{-E_g(g)} \ee has many of the same properties
as the original one (\ref{Zheuristic}). That is, it is assumed
that expectation values $\langle \mathcal{O}\rangle_\phi$ and
$\langle \mathcal{O} \rangle_{\phi=0}$ of a geometric observable
$\mathcal{O}$ computed in the two models have similar properties.
It is this assumption that motivates the study of
``pure-geometry'' systems even when the real universe is known to
contain matter fields of various types.

The purpose of this paper is to demonstrate that geometric
observables can in fact be different depending on whether or not
matter is included in the calculation. The strategy to demonstrate
this will be to compute expectation values explicitly in the
setting of a two dimensional rectangular box in which the sides
lengths $L_x$ and $L_y$ are variable but the total area $L_x
L_y=A$ is fixed. This system was already introduced in
\cite{Konopka:2008ds} but it is explored here more fully. An
important property of this system is that it contains a single
configuration where $L_x =L_y$ and the box is a square but it
contains a multitude of other configurations in which $L_x \neq
L_y$; the geometric observable $R=L_x/L_y$ differentiates the
square configuration from the non-square ones. The main result
will be that the expectation value for this ratio in a system
without matter must be different from one, but that it can be one
when matter is present.

The rest of the paper is organized as follows. Sec.
\ref{s_toygeometries} describes the flexible box system and
defines its sum over geometries. The expectation value of the
observable in the absence of matter is also computed. Sec.
\ref{s_particles} discusses the effect of adding particles. It is
shown through numerical studies that the expectation value of the
observable $R=L_x/L_y$ in the presence of particles can be close
to unity. This indicates that matter can select a homogenous and
isotropic geometry from the ensemble. In Sec. \ref{s_casimir} this
discussion is extended to the case of fields and it shown in what
circumstances the Casimir energy of fields has a similar effect. A
discussion of applications to current research programs on
``atomistic'' models\footnote{The term ``atomistic'' model is due
to H-T. Elze.} of spacetime are presented in Sec.
\ref{s_discussion}.

\section{Toy Dynamical Geometries \label{s_toygeometries}}

\subsection{Regularized Sum Over Geometries}

Formal expressions like (\ref{Zheuristic}) and (\ref{Zheuristic2})
do not specify the space of geometries to be integrated over or
how exactly to carry out the integrals. To overcome this, several
approaches (e.g. \cite{CDT,Konopka:2008hp,Konopka:2008ds}) propose
to study systems like (\ref{Zheuristic}) in a discretized setting.
In those approaches, one replaces the integral over geometries by
a finite sum over a set $\mathcal{C}$ of geometry configurations.
The same strategy is adopted here. Since the purpose of this paper
is to study effects of matter on dynamical geometries and not to
propose a sensible way or regularizing the path integral for
quantum gravity, the set $\mathcal{C}$ is chosen to make this
specific application possible.

Consider configurations spaces $\mathcal{C}_A$ corresponding to
sets describing rectangular boxes having sides of length $L_x$ and
$L_y$ and area $A$. To make each of these sets $\mathcal{C}_A$
finite, the possible lengths $L_x$ and $L_y$ must be restricted to
a discrete or semi-discrete spectrum characterized by at least one
length scale, $\ell$, designating a smallest allowed length. If
the allowed lengths are assumed to be equally spaced when they are
small, a concrete length spectrum that can be used is \be
\label{Lspectrum}
L \in \left\{ \begin{array}{ll} \mathbbm{N}\ell &\qquad{\mbox{if }}\, L \leq \lambda \\
\mathbbm{R}&\qquad {\mbox{if }}\, L > \lambda. \end{array} \right.
\ee The minimal value and next larger values for $L$ are $\ell$
and integer multiples of $\ell$, respectively, up to a scale
$\lambda$. Above this scale, $L$ can take any real number value.
It will be convenient to choose $\lambda = \sqrt{A}.$ With this
convention, the sets $\mathcal{C}_A$ can be written out explicitly
as \be \label{partitionsA} \mathcal{C}_A =\left\{ \,\left(\ell,
\frac{A}{\ell}\right), \, \left(2\ell,\frac{A}{2\ell}\right),\,
\ldots, \left( \sqrt{A},\sqrt{A}\right),\ldots, \,
\left(\frac{A}{\ell}, \ell\right) \right\}. \ee Note that these
sets are countable and well defined for all real $A>\ell^2.$

\subsection{Partition Function and Geometric Observables}

After defining the configuration space of a dynamical box with
area $A$, one must specify how the configurations should be
weighted in the ensemble defining the partition function
(\ref{Zheuristic}). In the case where they are all equally
weighted, the partition function of the system is \be
\label{Zdiscrete} Z = \sum_{p=1}^{\lambda/\ell} \left( 1+1\right)
= \frac{2\lambda}{\ell}=\frac{2\sqrt{A}}{\ell}\,. \ee The first
term shown in the sum represents values of $L_x$ in the discrete
part of the length spectrum, i.e. the first half of the
configurations listed in (\ref{partitionsA}). The second term is
the analogous sum over discrete values of $L_y$\footnote{The given
prescription for evaluating $Z$ carries a risk of over-counting
whenever the two configurations corresponding to $p=\lambda/\ell$
are in fact equivalent. This problem can be averted by considering
$A$ such that $\lambda/\ell$ is not an integer. In any case, the
error involved becomes negligible when the area is large.}. In the
end, $Z$ turns out to be equal to the size of the set
$\mathcal{C}_A$; the thermodynamic properties of the empty box
system are determined entirely by entropy.

Expectation values of observables describing the box geometry can
now be computed. For concreteness, consider the ratio $R =
L_x/L_y.$ The expectation value for $R$ should be close to unity
if the box is likely to be square, and very different from unity
if the box is likely to be long and narrow. One finds \be
\label{L22} \langle R \rangle = \frac{1}{AZ} \left[
\sum_{p=1}^{\lambda/\ell} \left(
(p\ell)^2+\left(\frac{A}{p\ell}\right)^{\!\!2\,}\right) \right]
\ra\frac{\pi^2 \sqrt{A} }{12 \,\ell }\,. \ee The last expression
denotes the evaluation of $\langle R \rangle$ in the limit
$A\ra\infty$. Since the result is divergent in this limit, (also
when $\ell\ra 0$), the calculation suggests that the box should be
expected to be long and narrow in that limit. Curiously, note that
the expectation value for the inverse ratio, $\langle L_y/L_x
\rangle$, is given by exactly the same diverging expression; this
is a consequence of the $L_x \leftrightarrow L_y$ symmetry of the
system and implies that these ratios provide measures for the
degree to which the two sides of the box are different and cannot
be used to infer which length, $L_x$ or $L_y$, is larger.

Another useful expectation value is \be \label{expR2} \langle R^2
\rangle = \frac{1}{A^2Z} \left[\sum_{p=1}^{\lambda/\ell} \left(
(p\ell)^4+\left(\frac{A}{p\ell} \right)^{\!\!4\,}\right) \right]
\ra \frac{\; \pi^4 A^{3/2}}{180\,\ell^3}; \ee the last expression
again stands for the $A\ra\infty$ limit. This quantity can be used
to compute the fluctuations $(\Delta R)^2 = \langle R^2\rangle -
\langle R\rangle^2$ of the ratio $R$. In the limit $A\ra \infty$,
the contribution to $(\Delta R)^2$ from the first term, $\langle
R^2 \rangle$, is much larger than that from the second, $\langle R
\rangle^2 $, and thus the fluctuations diverge as $A^{3/2}$.

\section{Particles in Dynamical Geometries \label{s_particles}}

\subsection{Particles as Standing Waves}

Suppose that the flexible box is populated with quantum mechanical
particles of mass $m$. The dynamics of each particle is described
by the Schrodinger equation in two dimensions with a potential
that is zero inside the box and infinity everywhere else. After
setting $\hbar = c = 1$, the energy spectrum for each particle is
\be \label{Espectrum} E = \frac{1}{2m} \left( k_x^2 + k_y^2
\right), \ee where $k_x$ and $k_y$ are the momenta in the $x$ and
$y$ directions, respectively. The allowed values for these momenta
($i=x,y$) are \be \label{allowedki} k_i = \frac{(2-\sigma)\pi
n_i}{L_i} \ee where $\sigma$ depends on the type of boundary
conditions, \be \label{allowedsigma}
\sigma =\left\{ \begin{array}{ll} 1&\qquad{\mbox{Dirichlet} }\\
0&\qquad \mbox{Periodic,} \end{array} \right. \ee and $n_i=\sigma,
1, 2, \ldots$ are quantum numbers whose range also depends on the
boundary conditions.

For a given box configuration $(L_x,L_y)$, the partition function
for a single particle is \be \label{littlezdef} z(L_x,L_y) =  \,
\sum_{n_x=\sigma}^\infty \mathrm{exp}\!\left({-\frac{\beta
(2-\sigma)^2 \pi^2 n_x^2}{2mL_x^2}}\right) \;
\sum_{n_y=\sigma}^{\infty} \mathrm{exp}\!\left({-\frac{\beta
(2-\sigma)^2 \pi^2 n_y^2}{2mL_y^2}}\right) \ee where $\beta=1/kT$
is the inverse temperature. For a system of $N$ identical
Maxwell-Boltzmann particles, their combined contribution is \be
z_N(L_x,L_y) = \left(z(L_x,L_y)\right)^N/N!\ee The total partition
for the particles as well as the box can then be written as \be
\label{bigZdef} Z = 2 \sum_{p=1}^{\lambda/\ell}
\,z_N\!\left(p\ell,\frac{A}{p\ell}\right) \ee where the factor of
two in front has the same origin as in (\ref{Zdiscrete}).

It is important to emphasize here that $z$ and $z_N$ depend
explicitly on the lengths of the sides of the box and the
parameter $\sigma$. In particular, it can be shown that the
largest term in $Z$ is due to configurations with $L_x =L_y$ when
$\sigma=1$, or to configurations with $L_x \gg L_y$ and $L_y \gg
L_x $ when $\sigma=0$. These features form the basis for
understanding the results described next.

\subsection{Evaluation of Observables}

The expectation value for $R$ can now be computed, analogously to
(\ref{L22}), using \be \label{L2difficult} \langle R \rangle =
\frac{1}{ZA} \sum_{p=1}^{\lambda/\ell} \left(\!
(p\ell)^2+\left(\frac{A}{p\ell}\right)^{\!\!2\,} \right)\;
z_N\!\left(p\ell,\frac{A}{p\ell}\right). \ee This can be evaluated
numerically given certain values for the parameters. To make for a
realistic situation, the mass of the particles and the temperature
of the system should be much smaller than the inverse minimal
length, $m,kT \ll 1/\ell$. Thus $\beta(2-\sigma)^2\pi^2/8m \gg
1/\ell^2$. After fixing this ratio in terms of $\ell$, which can
then be set to unity without loss of generality, the remaining
free variables are $A$ and $N$. In the following, the number of
particles $N$ is expressed in the form $ N = N_0 + c A^\alpha $
with $N_0$ and $\alpha$ positive dimensionless constants, and $c$
a positive constant with dimension depending on the value of
$\alpha$.

\begin{figure}[tbp]
  \begin{center}
    \mbox{
      \subfigure[]{\includegraphics[scale=0.58]{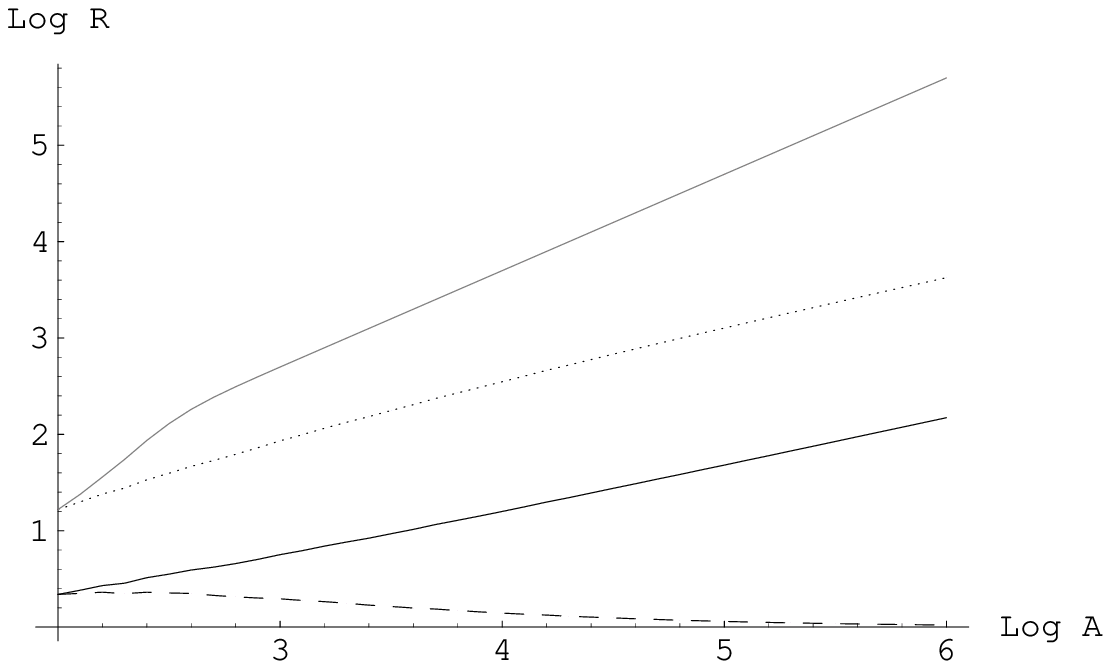}}
      \quad
      \subfigure[]{\includegraphics[scale=0.58]{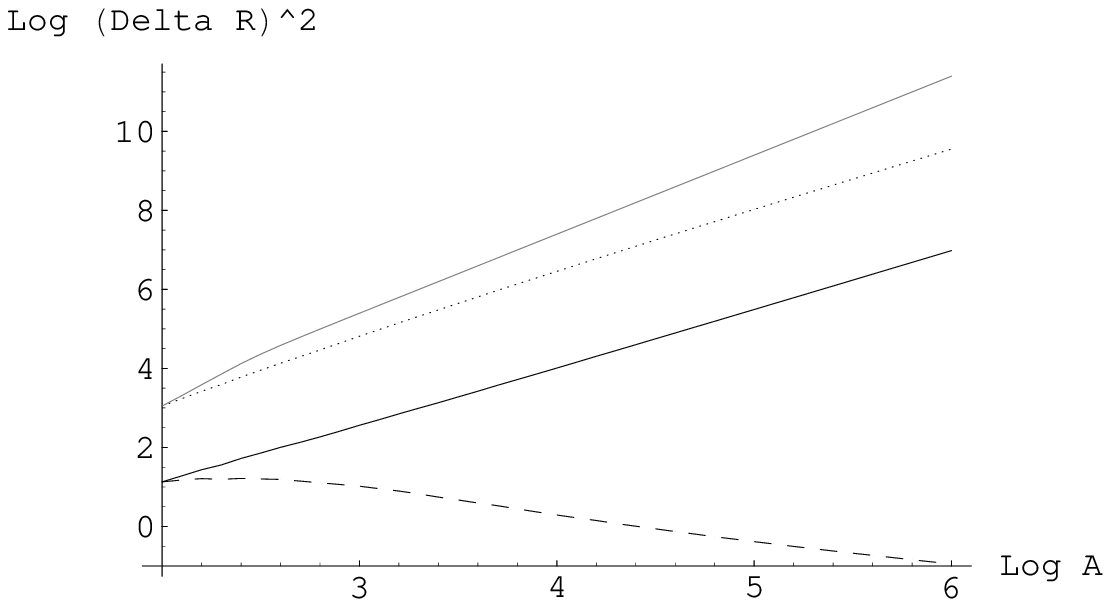}}
      }
    \caption{Geometric observables in toy geometries. In both plots, the lines
represent (top to bottom) systems with $(N,\sigma)=
(0.01A,0),\,(1,0),\,(1,1),\, (0.01A,1)$. }
    \label{fig_Nalpha}
  \end{center}
\end{figure}

Numerical results for $\langle R \rangle$ and $(\Delta R)^2$ with
$\ell=1$, $\beta(2-\sigma)^2\pi^2/8m = 10$ are shown in Fig.
\ref{fig_Nalpha}.  Curves corresponding to a fixed number of
particles ($N=1,\,\sigma=0,1$) show that the expectation value
$\langle R \rangle$ and its fluctuations rise as the area of the
boxes is increased. The asymptotic slopes match the behavior
expected from the empty box calculations in (\ref{L22}) and
(\ref{expR2}). Curves corresponding to systems wherein the number
of particles scales with area ($N=0.01A,\, \sigma=0,1$) exhibit
more interesting behavior. When Dirichlet boundary conditions are
imposed ($\sigma=1$), the asymptotic behavior for large $A$ is
$\langle R \rangle \ra 1$ and $\langle (\Delta R)^2 \rangle \ra
0$. Thus, in this case, the ensemble of particles mold the
dynamical geometry into a stable isotropic configuration. The
situation is different for periodic boundary conditions
($\sigma=0$); there both the expectation value for $R$ and its
fluctations diverge faster than in the empty box situations.

The most interesting result of these plots is the emergence of
isotropy, $\langle R \rangle \ra 1$, in the case of Dirichlet
boundary conditions. It can be understood qualitatively as
follows. The isotropic configuration contributes the most out of
all the terms composing the partition function $z(L_x,L_y)$ of a
single particle. This advantage is then raised to a power
$cA^\alpha$. Since the number of non-isotropic configurations
grows at most polynomially with $A$ (it grows as $\sqrt{A}$), all
of those configurations contribute negligibly to the overall
partition function in the large area limit. The difference in
growth rates between the polynomial and the exponential components
also explains why the approach to perfect isotropy occurs rather
quickly.

\section{Casimir Energy in Dynamical Geometries \label{s_casimir}}

\subsection{Regularization of Casimir Energy}

Another way of implementing matter is using fields instead of
particles. The expectation value of the Hamiltonian operator of a
field is in general given by a sum of two terms: one term
proportional to the number of particles in some state (zero for
the vacuum), and one other term that is divergent even in the
absence of physical particles. This latter term is called the
Casimir energy and it is usually attributed to quantum vacuum
fluctuations \cite{Bordag:2001qi}. It is (setting $\hbar=c=1$) \be
\label{CasimirEdef} E_C = \frac{\eta}{2} \sum_i \omega_i. \ee The
factor $\eta$ is positive for bosonic fields and negative for
fermionic fields. It is different from one if a field (such as a
vector field) can have multiple polarizations. The summation is
over all field mode energies.

In a flat $2+1$ dimensional background, relativistic field modes
have energies $\omega$ given by $\omega^2 = m^2 + k_x^2 + k_y^2$
in terms of momenta $k_{x,y}$ and mass $m$. The momenta are again
constrained by boundary conditions given in (\ref{allowedki}) and
(\ref{allowedsigma}). Thus the Casimir energy is \be
\label{CasimirEconc} E_0 = \frac{1}{2}\, \eta \pi (2-\sigma) \,
\sum_{n_x=\sigma}^\infty \sum_{n_y=\sigma}^\infty
\,\left(\widetilde{m}^2_\sigma+
\frac{n_x^2}{L_x^2}+\frac{n_y^2}{L_y^2}\right)^{\!\!1/2},\ee where
$\widetilde{m}_\sigma = m/(2-\sigma)\pi$. This quantity is
ultra-violet divergent; much work on the Casimir energy revolves
around extracting its physically relevant components
\cite{Bordag:2001qi}. One technique to do this is to define a
modified version of the summation (\ref{CasimirEconc}) which
includes a cutoff function $D_\ell$ multiplying each summand as
follows: \be \label{ECwithDell} E_{0,\ell} =
\frac{1}{2}\,\eta\sum_{n_x=\sigma}^\infty \sum_{n_y=\sigma}^\infty
\, \omega_{n_x,\,n_y} \,D_\ell\!\left(
\frac{n_x}{L_x},\frac{n_y}{L_y}\right). \ee The cutoff function
$D_\ell$ must be chosen to satisfy certain criteria
\cite{Cavalcanti:2003tw}. First, it must reduce to unity when
$\ell \ra 0$ so that $E_{0,\ell}\ra E_0$ in this limit. Second,
for $\ell\neq 0$, the function should fall off quickly when its
arguments become large as to make $E_{0,\ell}$ finite. Third, it
should not have any singularities or branch cuts in the complex
plane. Lastly, it is also convenient to choose $D_\ell$ so that it
is symmetric in its arguments, $n_x/L_x \leftrightarrow n_y/L_y$.

To evaluate (\ref{ECwithDell}), one applies the Abel-Plana formula
\be \sum_{n=0}^\infty F(n) = \int_0^\infty \!F(t) \, dt \,+\,
\frac{1}{2}F(0) \,+\, i \int_{0}^\infty \!\frac{dt}{e^{2\pi t}-1}
\left( F(it)-F(-it)\right) \ee twice in order to exchange the two
summations over $n_x$ and $n_y$ into integrals. The resultant
$E_{0,\ell}$ can be written (see
\cite{Cavalcanti:2003tw,Bordag:2001qi,Ambjorn:1981xw} for more
details) in the form \be \label{E0ellinffin} E_{0,\ell} =
\frac{1}{2} \, \eta \, \pi \,(2-\sigma) \,
(E_{\ell,\,A}+E_{\ell,\,B} + E_{AR}). \ee The first two terms in
the parenthesis are \be
\begin{split} E_{\ell,\,A} &= \left(L_x L_y\right)\, \int_0^\infty
\!du_x \int_0^\infty \!du_y \;
\sqrt{\widetilde{m}^2_\sigma+ u_x^2+u_y^2} \,\, D_\ell \left(u_x,u_y\right) \propto \, \frac{L_x L_y}{\ell^3} \\
E_{\ell,\,B} &= \frac{1}{2} \,(1-2\sigma) \left(L_x+L_y\right)
\int_0^\infty \!du\; \sqrt{\widetilde{m}^2_\sigma+u^2} \,\, D_\ell
\left(u,0\right) \; \propto \, (1-2\sigma)
\,\frac{L_x+L_y}{\ell^2}.
\end{split} \ee
These terms are divergent when $\ell\ra 0$ and thus parametrize
the ultra-violet behavior of $E_0$. When $\ell \neq 0$ and $m\ll
1$, their leading behavior is as shown. The precise coefficients
are dependent on the details of the cutoff function.
Interestingly, the terms are proportional to the area and
perimeter of the two-dimensional box, and the latter has different
sign for the two considered boundary conditions.

The last term in (\ref{E0ellinffin}), $E_{AR}$, consists of the
remaining contributions obtained from the expansions using the
Abel-Plana formula. For the present discussion, the important
features of $E_{AR}$ are that it is finite in the $\ell\ra 0$
limit and that its magnitude decreases with area
\cite{Bordag:2001qi,Milton:2004ya,Ambjorn:1981xw}.

The Casimir energy is an active subject of research
\cite{Bordag:2001qi,Milton:2004ya} because it is a purely quantum
effect that leads to a macroscopic effect in the form of the
Casimir effect/force. In the standard description of the Casimir
effect, it is the finite term $E_{AR}$ that is physically
interesting. A clear explanation of why the divergent terms can be
discarded when describing situations relevant for experiments
testing the Casimir force is given in works on cavities with a
piston \cite{Cavalcanti:2003tw}. In short, the reason is that the
divergent contributions from inside and outside a cavity cancel
from formulae describing measurable quantities (the force on the
piston).

In the context considered in this paper, the field is defined only
within the boundaries of the box, and thus a cancellation of
divergent terms from inside and outside cannot occur. The analysis
of the effect of Casimir energy on the statistical properties of
the dynamical box must therefore either include the regularized
divergent terms, or discard them in an ad-hoc manner.

\subsection{Casimir Energy and Expected Geometry}

The partition function of the flexible box with field is \be Z = 2
\sum_{p=1}^{\sqrt{A}/\ell} z_\phi(\beta, p) \ee where
$z_\phi(\beta,p)$ stands for the partition function of a field
theory at inverse temperature $\beta$ in a box configuration
labelled by $p$. $z_\phi(\beta,p)$ should have contributions from
the Casimir energy and also from the real particle present in a
thermal state. However, in the low temperature regime where the
average thermal energy is smaller than the mass of the particles,
it is reasonable to approximate $z_\phi$ by the contribution from
the Casimir energy alone. Thus $z_\phi(\beta,p) \sim e^{ -\beta
E_0(p)}$ with $E_0$ given by (\ref{E0ellinffin}) for each box
configuration labelled by $p$.

Consider first the option of including the divergent but
regularized terms of $E_0$ in the calculations. Since the area of
the boxes is kept fixed, the term $E_{\ell,A}$ contributing to
$E_0$ is the same for all the box configurations and so can be
factored out and essentially ignored. The other terms depend on
the geometry of the box and thus must all be taken into
consideration. Being divergent in the $\ell\ra 0$ limit, the term
$E_{\ell,B}$ proportional to the perimeter of the box is dominant
over the term $E_{AR}$. The latter term can therefore be ignored.
The partition function thus becomes \be Z =
2\sum_{p=1}^{\lambda/\ell} \; \mathrm{exp} \left(-\beta \eta\pi
(2-\sigma)(1-2\sigma)\frac{1}{\ell^2}\left(p\ell+\frac{A}{p\ell}\right)\right).
\ee Expectation values can be computed numerically similarly as
done in the previous section. The results are that the square
configuration become dominant if the parameters $\sigma$ and
$\eta$ are chosen as either $\sigma=1$ and $\eta<0$, or as
$\sigma=0$ and $\eta>0$. That is, the square configuration is
preferred if either the field is fermionic with Dirichlet boundary
conditions or if the field is bosonic with periodic boundary
conditions.

Next, consider the option of discarding the regularized divergent
terms of $E_0$ and using only the finite contribution $E_{AR}$.
Since $E_{AR}$ is decreasing with $A$, it's effect would be
insignificant in the partition function for the dynamical box. The
statistical properties of the dynamical box should thus be
expected to match those described in Sec. \ref{s_toygeometries}.
In other words, the finite component of the Casimir energy is not
sufficient for shaping the dynamical geometry into a particular
preferred configuration.

\section{Discussion \label{s_discussion}}

Any successful theory of quantum dynamical geometry must
eventually include coupling to the matter content observed in the
universe. A-priori, average properties of dynamical geometries
should be expected to be different when computed in theories with
and without matter. Consequently, it is important to understand in
detail the way in which matter may affect geometric observables.
This issue was addressed in this paper using a toy model in which
dynamical geometry was described by a two-dimensional ``flexible''
rectangular box of fixed area $A$. The expectation value of the
observable $R=L_x/L_y$ was shown to be very different depending on
whether matter was present or not, and depending on the boundary
conditions imposed on the matter. In particular, in the presence
of a finite density of particles and Dirichlet boundary
conditions, it was shown that $\langle R \rangle$ can approach
unity and the fluctuations $(\Delta R)^2$ can vanish in the large
area limit. This scenario can thus be argued to give rise to a
classical, stable, and isotropic geometry. A similar effect can
arise also in the case of quantum fields if the divergent but
regularized Casimir energy is included in the analysis.

The purpose of the calculations presented in this paper is to
exemplify {\em that} and illustrate {\em how} matter can affect a
dynamical geometry. One technical assumption that was used
regarded the form of the spectrum for the lengths $L_x$ and $L_y$.
Although the spectrum was chosen only for convenience and not
according to a deep physical principle, it should be possible to
formulate precisely the conditions that must be satisfied by the
physical length spectrum for the results presented to continue to
hold. This kind of analysis, however, goes beyond the scope of
this paper. Nonetheless, it is worth pointing out that the
spectrum used is a natural one from the point of view of
``atomistic'' models of spacetime models in which large-scale
geometry is thought of being composed of a large number of
elementary building blocks.

%\begin{figure}[tbp]
%  \begin{center}
%    \includegraphics[scale=0.28]{fig_2.eps}
%    \caption{Graphs generated by graphity models \cite{Konopka:2008ds}: a tree-like graph generated by a basic version of a graphity model, and a homogenous, lattice-like graph generated by another graphity model. }
%    \label{fig_graphs}
%  \end{center}
%\end{figure}

A natural application of the presented ideas is in the context of
``atomistic'' models of spacetime. In this context, conventional
approaches to studying matter on dynamical geometries (see e.g.
\cite{BirrellDavies}) have limited use because the notion of
geometry is only emergent, but the general mechanism described in
this paper can be applied. Consider for example the class of
models called graphity \cite{Konopka:2008ds,Konopka:2008hp}: the
basic version of these models does not generate extended
manifold-like graphs, but modified versions that include a
homogeneity requirement do \cite{Konopka:2008ds}. It has been
suggested that matter degrees of freedom might provide such a
homogeneity requirement via a mechanism similar to the one
described here. Similarly, it is possible that matter might play
an important role in other models of emergent spacetime such as
causal sets \cite{Rideout:2008rk,Johnston:2008za} or group field
theory \cite{Oriti:2007qd}.

Another application of the ideas presented here is in cosmology:
instead of putting homogeneity and isotropy as inputs in an ansatz
for a cosmological spacetime solution, it may be interesting to
see them arising from the statistical mechanics of matter fields.

\ack I would like to thank the organizers of DICE 2008 for a
diverse and extremely stimulating workshop. I have also benefitted
from discussions with J. Ambjorn and B. Z. Foster.


\medskip


\begin{thebibliography}{99}


\bibitem{DynamicalGeometry} Oriti D, ed. 2008 {\em Approaches
  to Quantum Gravity}, Cambridge University Press.

\bibitem{BirrellDavies} Birrell ND and Davies PC 1982 {\em Quantum
fields in curved space}, Cambridge University Press.

\bibitem{Konopka:2008ds}
  Konopka T,
  %``Statistical Mechanics of Graphity Models,''
  Phys.\ Rev.\ D {\bf 78}, 044032 (2008) [arXiv:0805.2283 [hep-th]].

\bibitem{Konopka:2008hp}
  Konopka T, Markopoulou F and Severini S,
  %``Quantum Graphity: a model of emergent locality,''
  Phys.\ Rev.\  D {\bf 77}, 104029 (2008)
  [arXiv:0801.0861 [hep-th]].

\bibitem{CDT}
  Ambjorn J, Jurkiewicz J and Loll R,
  %``A non-perturbative Lorentzian path integral for gravity,''
  Phys.\ Rev.\ Lett.\  {\bf 85}, 924 (2000).
  [arXiv:hep-th/0002050].

\bibitem{Milton:2004ya}
  Milton KA,
  %``The Casimir effect: Recent controversies and progress,''
  J.\ Phys.\ A  {\bf 37}, R209 (2004).
  [arXiv:hep-th/0406024].

\bibitem{Bordag:2001qi}
  Bordag M, Mohideen U and Mostepanenko VM,
  %``New developments in the Casimir effect,''
  Phys.\ Rept.\  {\bf 353}, 1 (2001)
  [arXiv:quant-ph/0106045].

\bibitem{Cavalcanti:2003tw}
  Cavalcanti RM,
  %``Casimir force on a piston,''
  Phys.\ Rev.\  D {\bf 69}, 065015 (2004)
  [arXiv:quant-ph/0310184].

\bibitem{Ambjorn:1981xw}
  Ambjorn J and Wolfram S,
  %``Properties Of The Vacuum. 1. Mechanical And Thermodynamic,''
  Annals Phys.\  {\bf 147}, 1 (1983).

%\bibitem{Henson:2006kf}
%  J.~Henson,
%  %``The causal set approach to quantum gravity,''
%  arXiv:gr-qc/0601121.

\bibitem{Rideout:2008rk}
  Rideout D and Wallden P,
  %``Spacelike distance from discrete causal order,''
  arXiv:0810.1768 [gr-qc].

\bibitem{Johnston:2008za}
  Johnston S,
  %``Particle propagators on discrete spacetime,''
  Class.\ Quant.\ Grav.\  {\bf 25}, 202001 (2008)
  [arXiv:0806.3083 [hep-th]].

\bibitem{Oriti:2007qd}
  Oriti D,
  %``Group field theory as the microscopic description of the quantum
  %spacetime fluid: a new perspective on the continuum in quantum gravity,''
  arXiv:0710.3276 [gr-qc].




\end{thebibliography}
\end{document}